\documentstyle{l-aa}

\def\M0b{\overline{M}_0}
\def\disp{\displaystyle}

\def\d{{\rm d}}
\def\mR{{\cal R}}
\def\mC{{\cal C}}
\def\mA{{\cal A}}
\def\mD{{\cal D}}

\def\Id{{\rm Id}}

\def\vg{{\vec g}}
\def\vx{{\vec x}}
\def\vk{{\vec k}}
\def\vl{{\vec l}}
\def\ii{{\rm i}}

\def\gam{{\vec{\gamma}}}

\def\mg{\big <}
\def\md{\big >}

\def\ort{\perp}

\def\delT{\delta_T}
\def\dTo{\delta_T^{\rm obs.}}
\def\dTp{\delta_T^{\rm prim.}}
\def\chicmb{\chi_{\rm CMB}}
\def\be{\begin{equation}}
\def\ee{\end{equation}}
\def\ba{\begin{eqnarray}}
\def\ea{\end{eqnarray}}
\def\p2d{p_{\rm 2d}}
\def\Npk{N_{\rm peak}}
\def\npk{n_{\rm peak}}
\def\Prob{{\rm Pr.}}

\input psfig.sty

\begin{document}

   \thesaurus{12 (12.03.1; 12.04.1; 12.07.1; 12.12.1)} 

 \title{Lens Distortion Effects on CMB Maps}

 \author{F. Bernardeau}

 \offprints{F. Bernardeau; fbernardeau@cea.fr}

 \institute{Service de Physique Th\'eorique, 
C.E. de Saclay, F-91191 Gif-sur-Yvette cedex, France\\}

\maketitle

\markboth{Weak Lensing on CMB Maps}{F. Bernardeau}

\begin{abstract}

Weak lensing effects are known to introduce non-linear couplings
in the CMB temperature maps. In inflationary scenario, the primary CMB 
anisotropies are expected to form a 2D Gaussian map, for which, the
probability distribution function of the ellipticity defined
from the local temperature curvature matrix has a very
specific shape. I show that lenses alter significantly the shape
of this PDF, inducing an excess of elongated structures. The precise
functional form is computed for both the field points and the
temperature extrema. 

These analytical results are confirmed by numerical experiments
on 10x10 square degree maps. These numerical results allow to investigate
the effects of smoothing and to estimate the cosmic variance. 
For the best resolution and sky coverage of the Planck mission the
signal to noise ratio for the statistical indicators presented here
is about 3 to 6 depending on the cosmological models. A marginal
detection should therefore be possible.


\end{abstract}

\section{Introduction}
The detection by the COBE/DMR experiment (Smoot et al. 1993) of the
anisotropies of the CMB temperature at very large angular scales,
above $7\deg$, 
has open an exiting new mean of investigation for cosmology. Many
experiments that are now under development will provide us in the
near future with precious data at smaller angular scale, down to a 
fraction of a degree.  At such angular scales the dominant mechanism that
generates the CMB anisotropies is the primary, linear order, coupling
of the various cosmic fluids (photons, baryons and
different species of possible dark matter) before and during
recombinaison. Whether the primary anisotropies originate from 
quantum fluctuations in an inflationary scenario, or from other
mechanisms such as topological defects generated in a phase transition
epoch is yet unclear. This is undoubtlessly a major scientific goal for the
coming experiments. Indeed, one of the clearest signature of the topological
models is that they are expected to
induce non-Gaussian primary temperature fluctuations
(Pen et al. 1994, Turok 1996, Barnes \& Turok 1996). 
But even in the case of inflationary
scenario, it is possible that secondary effects, non-linear couplings
of the radiation field with matter, induce non-Gaussian
features. This is particularly
important at very small angular scale, down to the arcmin scale. 
I am more particularly interested here in the static
lens effects. The effects of the gravitational distortion on
the power spectrum of the primary temperature maps have been 
the subject of many investigations in the last decade
(Blanchard \& Schneider 1987, Kashlinsky 1988,
Cole \& Efstathiou 1989, Sasaki 1989, Tomita \& Watanabe 1989, 
Linder 1990, Cay\'on, Mart\'\i nez-Gonz\'alez \& Sanz 1993a, b,
Fukushige, Makino \& Ebisuzaki 1994, Seljak 1996).
It is now clear that the lens effect on the power spectrum is small,
although it is probably worth to take it into account
in a very detailed analysis of such measurements.
In principle, for an inflationary scenario, a detailed analysis
of the CMB power spectrum should allow to constrain very accurately
the cosmological parameters (see for instance Jungman et al. 1996).
But it would be anyway extremely interesting to be able to have
a positive detection of the lens effect since it can potentially
be used to constrain the amplitude of the cosmic density 
power spectrum, independently of the constraints
obtained from the CMB power spectrum itself. It could therefore
be a precious test for the cosmological
model(s) favored by the shape of the CMB power spectrum.

In a previous paper, the four point functions
induced by the lens effects has been investigated
(Bernardeau 1997). This is the most direct quantity that can be 
calculated from the coupling terms introduced by the lenses.
One thus obtains specific properties of CMB maps induced by the lenses.
Another possible way of investigation is to look
for cross-correlation between the CMB temperature gradients and
the displacement field induced by the projected large-scale structures
(Suginohara et al. 1997). The signal however depends on the bias properties
of the galaxies. It cannot be directly interpreted in terms of cosmological
parameters.

If the four point function is indeed sensitive to the
intrinsic depth of the lens potential wells, it
is not necessarily the best indicator in terms of signal to noise
ratio for a realistic experiment.
The aim of this paper is therefore to investigate new
means of detecting these effects. In section 2 I present 
numerical results showing peculiar examples of the lens effects on
temperature maps. The visual impression is that the
lens effects induce a change in the topological properties of the 
temperature maps, and not so much in the local temperature distribution
function.

This is the motivation for the investigation of other 
statistical indicators that can reveal the lens effects with a
better efficiency.
Particularly interesting can be the statistical properties of the
local curvature of the temperature maps.
The matrix of the second order temperature derivatives 
has indeed specific statistical properties in case
of a 2D Gaussian field. These properties are affected when the
lens effects are taken into account because of the induced mode couplings. 
More specifically a quantity that can be a good tracer of the
lens effect is the probability distribution function of the local
ellipticity since lenses tend to systematically 
stretch the local temperature patches. Compared to the 4-point
function the motivation for considering such a quantity is then double.
First of all, the cosmic variance is expected to be smaller: It is given
by statistical properties related to the second order derivative, and
therefore more sensitive to the small scale temperature fluctuations.
Secondly the effects of lenses on the four-point function were 
found to be proportional to the cosine of the angle joining the
observational directions (see Bernardeau 1997), which tends to substantially
reduce the lens signal when it is averaged for the computation
of the four point moment. Such a cancellation is not expected for the
distortion effect on the curvature. The derivation of the lens effect for the
distribution function of the local ellipticity is presented
in Section 3. In Section 4, results are confronted with numerical
experiments. In particular I estimate the cosmic variance for the
ellipticity statistics. In the last section I discuss the dependence
of the lens effects on the cosmological parameters.

\section{The physical mechanisms}

First of all let me recall the basic effect induced by the lenses.  I
am interested here in the static lenses caused by the presence along
the line-of-sights of large-scale concentrations of matter. These
concentrations of matter are of course intrinsic to any theory of
large-scale structure formation. The effects presented in this paper
are therefore robust consequences of the growth of structures.  The
amplitude of the effects, however, depends not only on the amplitude
of the local fluctuations but also on the global cosmological
parameters: the lens effects are all the more large that the CMB plane
is distant from the lenses. The dependence
of the signal on the cosmological parameters is discussed in the
last section.  The static lenses (contrary to the case of moving lenses, see
Birkinshaw \& Gull 1983) do not change the background temperature of
the CMB, but only induce a displacement field so that the observed
temperature fluctuation in a given direction $\gam$ is the one coming
from a slightly shifted direction, $\gam+\xi(\gam)$, 
\be
\dTo(\gam)=\dTp(\gam+\xi(\gam)), \label{leff}
\ee 
where $\xi(\gam)$ is the
displacement field induced by the lenses along the line of sight in
the direction $\gam$. Written in terms of the gravitational potential $\Psi$,
the displacement reads, 
\be 
\xi(\gam)=-2\int_0^{\chi_{\rm CMB}} \d\chi\
{\mD(\chi_{\rm CMB},\chi)\over \mD(\chi_{\rm
CMB})}\nabla\psi(\gam,\chi),
\ee 
where $\chi$ and $\mD$ are
respectively the radial and angular distances.  These quantities 
are identical in
case of a background universe with zero  curvature. $\mD_{\rm CMB}$ is
the angular distance to the last scattering surface. The lens
population identifies with the large-scale structures of the Universe that
are present along the line of sights. It is important to have in mind
that the displacements induced by these matter concentrations are at most
of the order of 1 arcmin for cluster cores, whereas the instrumental 
resolution for the future satellite missions\footnote{Note however that
ground based interferometer instruments can potentially reach a better 
resolution.} is at best 5 arcmin. 
As a result, it is safe to neglect the influence of the
critical regions and the lens effect
is assumed here to be fully in the weak lensing regime. 

A consequence of this remark is that the effects investigated in this paper
can be entirely described by the two-point correlation 
properties of the lens population.
The precise reason is that the displacement field being a small
perturbation of the primordial field, its leading
contribution can be expressed perturbatively solely
with the two-point correlation function of either
the displacement field or its derivatives such as the local gravitational
convergence. This situation was already encountered in the
computation of the induced temperature four-point function
(Bernardeau 1997).
Therefore, the lens effects will depend quantitatively only on
the shape and magnitude of the cosmic density 
power spectrum, $P(k)$, defined
from the Fourier transform of the local density field, 
\be
\delta(\vx)=\int{\d^3 \vk\over (2\pi)^{3/2}}\,
\delta_{\vk}\,\exp(\ii\vk\cdot\vx), \ee 
with then, 
\be
\mg\delta_{\vk}\delta_{\vk'}\md=\delta_{\rm Dirac}(\vk+\vk')\,P(k).
\ee 
Note that in this analysis, $P(k)$, is the actual power spectrum,
that includes possible nonlinear evolution of the density field.
Written in terms of the modes $\delta_{\vk}$ the displacement field reads, 
\ba 
\xi(\gam)&=\disp{
\int_0^{\chicmb}\d\chi\ w(\chi)\int\disp{\d^3\vk\over (2\,\pi)^{3/2}}}
\times\label{dgam}\\ &\disp{\ii\,\vk_{\ort}\over k^2\,\mD(\chi)}\
\delta(\vk)\, \exp\left[ \ii \mD(\chi)\,\vk_{\ort}\cdot\gam+\ii
k_r\,\chi\right],\nonumber 
\ea 
with, 
\be w(\chi)=\disp{{3\Omega_0\over a(\chi)}}\,
\disp{\mD(\chicmb-\chi)\,\mD(\chi)\over\mD(\chicmb)}.\label{eff} 
\ee 
The function $w(\chi)$ gives the lens efficiency function
for sources located on the last scattering surface.  The second moment
of the displacement field  then reads 
\ba &\mg\xi^2\md=\disp{
\int_0^{\chicmb}\d\chi\ w(\chi)\,\int_0^{\chicmb}\d\chi'\ w(\chi')
\int\disp{\d^3\vk\over (2\,\pi)^{3}}}\times\nonumber\\
&P(k)\,\disp{\vk_{\ort}^2\over k^4\mD^2}\, \exp\big[\ii
(\mD(\chi)\!-\!\mD(\chi')\,\vk_{\ort}\!\cdot\!\gam+ \ii
k_r\,(\chi\!-\!\chi')\big].  
\ea 
Applying the small angle
approximation, the integral in $k_r$ introduces a Dirac function in
$\chi=\chi'$ so that  
\ba \mg\xi^2\md&=\disp{ \int_0^{\chicmb}\d\chi\
w^2(\chi) \int\disp{\d^2\vk\over (2\,\pi)^{2}}\  \disp{P(k)\over
k^2\mD^2}}\,.  
\ea 
It is then simple to see that the displacement
field can be expressed in terms of a 2d projected potential, $p_{\rm
2d}$, defined by 
\be 
\p2d(l)=\disp{\int_0^{\chicmb}\d\chi\
\disp{w^2(\chi)\over\mD^2} P\left[{l/\mD(\chi)}\right]}, \ee with \be
\mg\xi^2\md=\disp{ \int\disp{\d^2\vl\over (2\,\pi)^{2}}\
\disp{\p2d(l)\over l^2}}\,.  
\ee 
The lens effects are thus entirely
determined by $\p2d$.

\subsection{Numerical Experiments}

\begin{figure*}
\vspace{10 cm}
\special{hscale=60 vscale=60 voffset=-120 hoffset=-50 psfile=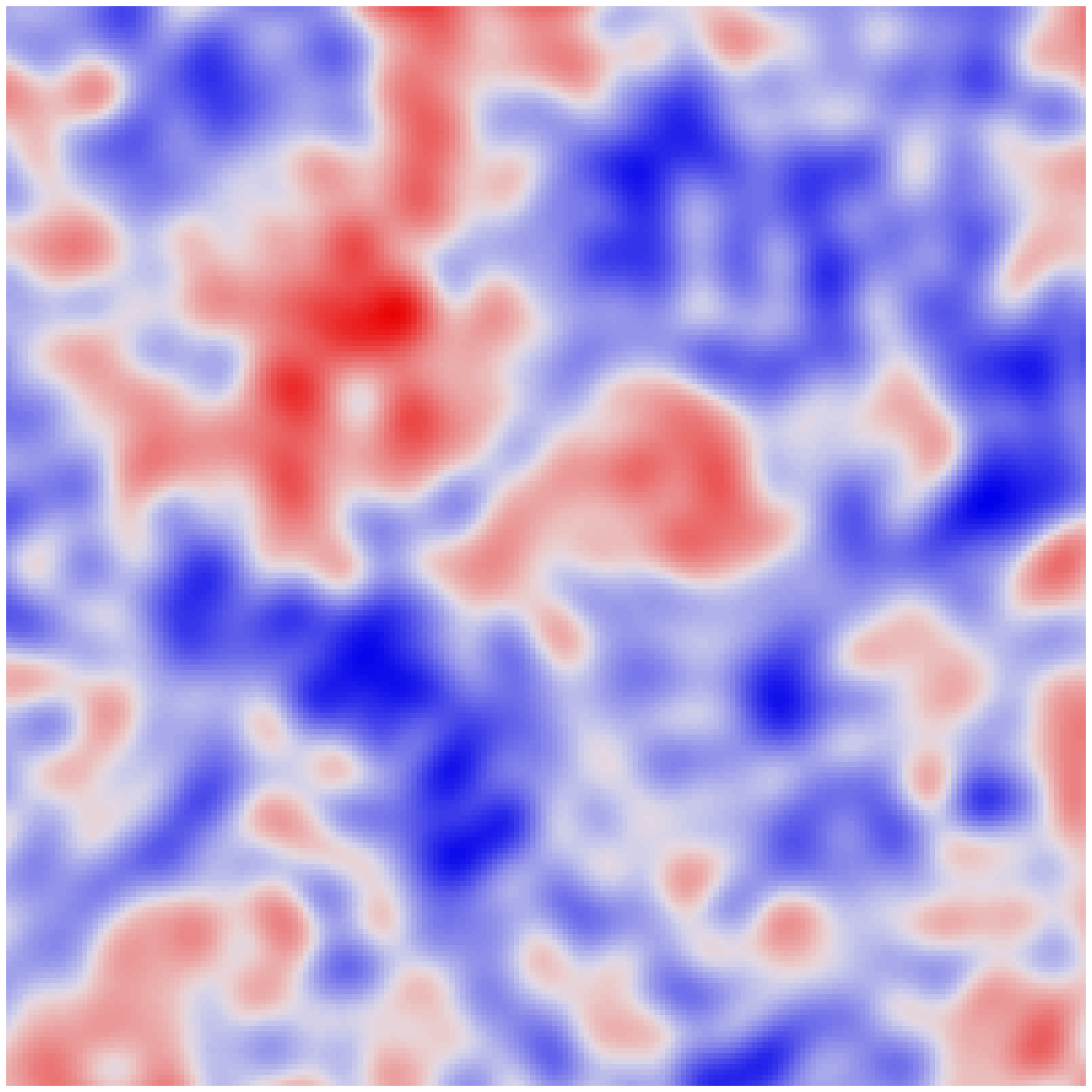}
\special{hscale=60 vscale=60 voffset=-120 hoffset=200 psfile=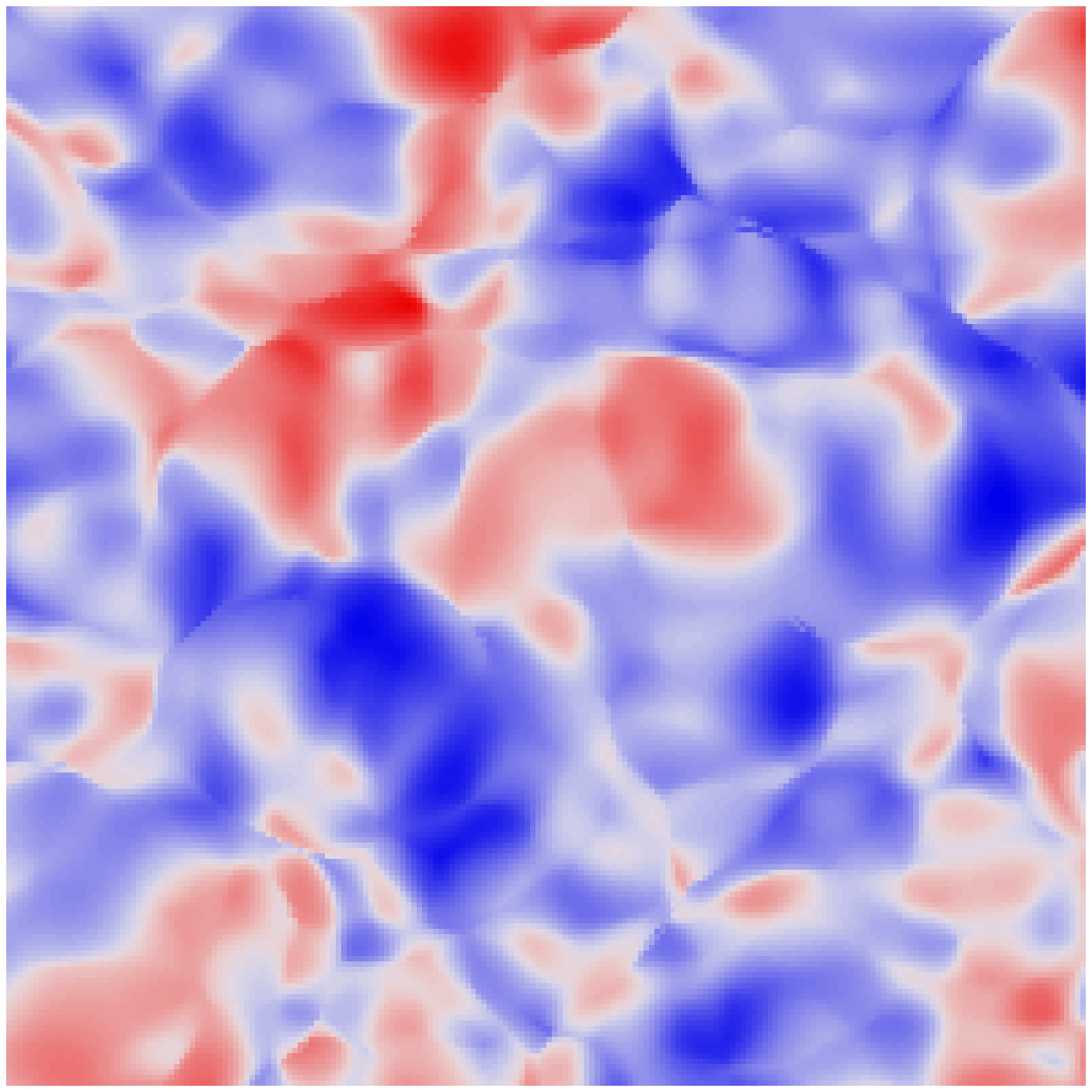}
\caption{An example of primordial temperature map (left)
deformed by the lens effect (right). It corresponds to a CDM
model. The figures are 4x4 square degree large. The lensed effects
have been magnified to make them visible. The displacement field
has been multiplied by about 4. }
\end{figure*}

The effect of the mechanism described in (\ref{leff}) can be easily
visualized in a simple numerical experiment. A 
realistic temperature map can be generated from a given
power spectrum, $C_l$. A 2D displacement
field can also be generated 
as a realization of a Gaussian process with the power
spectrum $\p2d$ obtained from a given 3D power spectrum. 
In the following I use a standard CDM model
for both the temperature map and the displacement field. 
The normalization of the temperature power spectrum is arbitrary;
the one of the displacement field will be discussed in terms of
$\sigma_8$, i.e. the r.m.s. of the density fluctuations at
$8 h^{-1}$ Mpc scale. A technical simplification is to assume that
the displacement field is Gaussian. This is obviously not exact,
specially at small scale,
but as it has already been mentioned, the lens effects depend
only on the power spectrum of the projected lens density.
To mimic the effects of the lens on the CMB maps it is thus
not necessary to have a full 3D realization of the large-scale
structures. Note that 
the non-linear effects for the evolution of the lens density
field are implicitly taken into account with the use of the non-linear power
spectrum $P(k)$ (Jain et al. 1995, Peacock \& Dodds 1996,
following the approach of Hamilton et al. 1991). The effects
of the lenses are then computed from the equation (1): it means
that the lensed temperature map is given by the primordial one,
but on a now irregular grid. The lensed temperature map is obtained
by a ``regridding'' of the temperature map, with a local linear interpolation
of the temperature with the nearest neighbors. This algorithm is
usually available in standard mathematical packages.

The two maps of Fig 1 show the effects of lenses. The effect is
of course magnified so that it can be easily detected by a visual inspection
of a very small map. The value of $\sigma_8$ would be here of about 2.
The size of the maps is 4x4 degrees, for a pixel size of 1.17'.
On the lensed map, one can see that the temperature patches are still
present with the same temperature contrasts. They are however 
displaced and deformed. The most striking feature is
probably that there are very sharp edges and
extended area of constant gradients. All
these features suggest that the local curvature is probably a good 
indicator of the lens effects.

\section{The lens effects on the local temperature curvature}

\subsection{The local curvature for the primordial temperature map}

The temperature curvature is defined from the matrix of the second
order derivatives of the local temperature,
\ba
c_{ij}\equiv&\disp{\d^2\delT\over \d x_i \d x_j}
\ea 
that can be written,
\ba
c_{ij}=&\left(
\begin{tabular}{cc}
\vspace{.2cm}
${\d^2\delT\over (\d x_1)^2}$ & ${\d^2\delT\over \d x_1 \d x_2}$ \\
${\d^2\delT\over \d x_1 \d x_2}$ & ${\d^2\delT\over (\d x_2)^2}$
\end{tabular}
\right) \equiv \left(
\begin{tabular}{cc}
$\tau+g_1$ & $g_2$ \\
$g_2$ & $\tau-g_1$
\end{tabular}
\right).
\ea
The previous equation defines the scalar field $\tau$ and the 
vector field $\vg$. Note that a rotation of the coordinate system
with an angle $\theta$ transforms the components of $c_{ij}$ in 
\ba
\tau'&\to\tau,\nonumber\\
\vg'&\to \mR_{2\theta}\cdot\vg,\nonumber\\
\ea
where $\mR_{2\theta}$ is the rotation matrix of angle $2\theta$.
Thus the vector $\vg$ behaves rather like a pseudo-vector.
For a Gaussian temperature field it is easy to show that
$\tau$, $g_1$ and $g_2$ are three Gaussian {\it independent} fields. 
At a given point their variances are given by,
\ba
\mg\tau^2\md=&\sigma_2^2,\\
\mg g_1^2\md=&\mg g_2^2\md={1\over 2}\sigma_2^2,
\ea
with $\sigma_2$ given, in the small angle approximation, by
\be
\sigma_2^2={1\over 4}\int{\d^2\vl\over (2 \pi)^2}\ l^4\ C_l,
\ee
where $C_l$ is the temperature power spectrum.
This entirely defines the statistical properties of the quantities
associated with the primordial temperature field. 

A quantity of interest is then the ellipticity,
\be
e={g\over 2 \tau},
\ee
where $g$ is naturally given by,
\be
g=\sqrt{g_1^2+g_2^2}.
\ee
The local ellipticity expresses the fact the temperature
fluctuations may be locally isotropic or not. Let me be a bit more precise.
First of all it is always possible to assume that $g_2$ equals
zero (by a proper choice of coordinates) so that $g=\vert g_1\vert$. 
Consider then the curvature matrix $c_{ij}$ and its 2 eigenvalues.
They identify with $\tau-g$ and $\tau+g$. If they have
the same sign, as it is the case for a local extremum,
it is also the sign of $\tau$ and consequently the one of the ellipticity. 
In such a case, when the ellipticity is positive, $\tau -g$ is positive and
the ellipticity is smaller than $1/2$. In general 
the absolute value of extremum ellipticities is less than 1/2, i.e.
elongated peaks correspond to 
ellipticities close to $1/2$ and round structures close to 0
whereas values larger than $1/2$ correspond to regions
of eigenvalues of different sign such as saddle points\footnote{Note that
although this discussion
is focussed on points of vanishing temperature gradient, the curvature
tensor and the local ellipticity are defined on every field point.}.

The distribution function of the local ellipticity can be easily 
computed from the known statistical properties of $\tau$
and $g$. One finds that (eg. Bond \& Efstathiou 1987),
\be
p_e^{\rm prim.}(e)\d e= {4\, \vert e\vert
\,\d e\over (1+8\, e^2)^{3/2}}.\label{eqpe}
\ee
As there are no statistical differences at all between the hot and the
cold regions in the primordial map, and since the lens effects do
preserve this property, in the following I will 
consider only the cases of positive ellipticities. 
It is important to note that {\it the distribution (\ref{eqpe})
does not depend on the shape of the temperature power spectrum}.
It is specific of a 2D Gaussian field. 

In (\ref{eqpe}) $e$ runs from 0 to $\infty$ (when limited to a positive
ellipticity).
For practical reason I make the change of variable,
\be
\epsilon=1/\left({1\over 2\,e}+1\right)
\ee
so that $\epsilon$ runs from 0 to 1. The distribution of $\epsilon$
can be easily computed from the one in $e$,
\be
p_{\epsilon}^{\rm prim.}(\epsilon)\d\epsilon={2\,\epsilon\,\d\epsilon\over 
(1-2\,\epsilon+3\,\epsilon^2)^{3/2}},\ \ \ 0\le\epsilon\le1.
\ee
This distribution peaks at $\epsilon=0.5$.
The numerical results will be discussed in terms of $\epsilon$
because a binning of constant width is obviously more
appropriate for $\epsilon$ compared to $e$.

The effect of lenses is, from the visual inspection of figure 1, to
extend the regions with very asymmetric curvature, that is to favor regions
of ellipticity close to $1/2$. The aim of the coming sections is to quantify 
analytically this effect.

\subsection{The lens effects}

\subsubsection{The lens effects on the curvature matrix}

We have now to consider the local curvature matrix 
in the presence of lenses. The first derivative of eq. (\ref{leff})
leads to
\ba
(\dTo)_{,i}=(\dTp)_{,j}(\delta^K_{ji}+\xi_{j,i})
\ea
where $\delta^K_{ij}$ is the Kronecker symbol. This expression makes
naturally intervene the lens amplification matrix, $\mA_{ij}$,
\be
\mA_{ij}\equiv \Id+\big(\xi_{i,j}\big)=\left(
\begin{tabular}{cc}
$1-\varphi_{,11}$&$-\varphi_{,12}$\\
$-\varphi_{,12}$&$1-\varphi_{,22}$
\end{tabular}
\right),
\ee
where $\Id$ is the identity matrix and
$\varphi$ is the projected potential of the matter density
fluctuations along the line-of-sight (the power spectrum of which 
obviously is $\p2d(l)/l^2$).
Note that in the weak lensing regime the matrix 
$\mA$ is always regular, so that the extrema are conserved
by the lens effects (at least as long as filtering is not
taken into account).

The second order derivatives of eq. (\ref{leff}) lead to
\ba
(\dTo)_{,ij}=&(\dTp)_{,kl}(\delta^K_{ki}+\xi_{k,i})(\delta^K_{lj}+\xi_{l,j})
+\nonumber\\
&(\dTp)_{,k}\ \xi_{k,ij}.
\ea
Two terms appear. The second term vanishes for a peak, and
in such a case the lens effect can be written in a rather compact way,
\be
\mC^{\rm obs.}=\mA\cdot\mC^{\rm prim.}\cdot\mA,\label{ACA}
\ee
where $\mC$ is the curvature matrix. This expression is familiar
in studies of weak lensing effects on background galaxies.
It is also interesting to note that a peak cannot be transformed in
a saddle point, and inversely: 
from relation (\ref{ACA}) it appears that the determinants
of $\mC^{\rm prim.}$ and $\mC^{\rm obs.}$ have the same sign, as long
as the lens effect is not critical.

\subsubsection{The local ellipticity distribution}

The computation of the statistical properties of the
matrix element of the curvature matrix of the lensed
map then requires the knowledge of the statistical
properties of the lens contributions. First of all the displacement field is
statistically independent of the temperature fluctuations of the
CMB. Secondly the displacement is small, so that it is safe
to do a perturbative calculation
with respect to the displacement. It implies that only the
second moment of the displacement, and of its derivatives, will appear
in the final results.

The components of the amplification matrix $\mA$ can be written
in terms of the local gravitational convergence $\kappa$ and
distortion $\gamma$,
\be
\mA\equiv\left(
\begin{tabular}{cc}
$1-\kappa-\gamma_1$ & $-\gamma_2$\\
$-\gamma_2$ & $1-\kappa-\gamma_1$ 
\end{tabular}
\right).
\ee
Similarly to the case of the curvature matrix the r.m.s.
values of $\kappa$, $\gamma_1$ and $\gamma_2$ are related to each other
through,
\be
\mg\gamma_1^2\md=\mg\gamma_2^2\md={1\over2}
\mg\kappa^2\md\equiv {1\over 2}\sigma_{\kappa}^2,
\ee
although $\kappa$ and $\gamma_i$ may not be Gaussian distributed.
This is a simple
consequence of geometrical averages, that take advantage of the assumed
statistical isotropy of the displacement field.
The second moment of the local convergence,
$\sigma_{\kappa}^2$, is related to the projected power spectrum 
with\footnote{This expression ignores the effects of filtering and is thus not
realistic. I reconsider this at the end of this section.},
\be
\sigma_{\kappa}^2=\disp{{1\over 4}\,
\int{\d^2\vl\over (2\pi)^2}\,\p2d(l)}.
\ee
The relation between $\sigma_{\kappa}$ and the normalization of the
local density spectrum, $\sigma_8$, makes intervene the
projection effects that depend on the cosmological parameters.
That will be discussed in more detail in the last section.

The second order derivatives are all independent of the first order 
derivatives. Only 4 different terms are generated, $\xi_{1,11}$,
$\xi_{1,12}$, $\xi_{1,22}$ and $\xi_{2,22}$. The others are identical
to one of those 4 because the displacement is potential.
Then we have,
\be
\begin{tabular}{lcl}
$\mg\xi_{1,11}^2\md$&=&${5/ 16}\ s^2$,\\
$\mg\xi_{2,22}^2\md$&=&${5/ 16}\ s^2$,\\
$\mg\xi_{1,12}^2\md$&=&${1/ 16}\ s^2$,\\
$\mg\xi_{1,22}^2\md$&=&${1/ 16}\ s^2$,\\
$\mg\xi_{1,11}\xi_{1,22}\md$&=&${1/ 16}\ s^2$,\\
$\mg\xi_{1,12}\xi_{2,22}\md$&=&${1/ 16}\ s^2$,
\end{tabular}
\ee
and the other cross-correlation terms are zero. This is again a simple
consequence of geometrical averages. The quantity $s$
is also related to the projected power spectrum,
\be
s^2=\disp{
\int{\d^2\vl\over (2\pi)^2}\,l^4\,\p2d(l)}.
\ee

The calculation of the statistical properties of the curvature
elements is then now quite straightforward. One has to (perturbatively)
invert the relation between the quantities associated with the
lensed map and the primordial map, and then to compute
the distribution function $\tau$ and $g_i$ from their
distribution function in the unlensed case.
These calculations are in practice a bit lengthy but can be performed
without much difficulty with a formal calculator.
One then has,
\be
p^{\rm obs.}(\tau,g_1,g_2)=
p^{\rm prim.}(\tau,g_1,g_2)\left[1+Q_1\,\sigma_{\kappa}^2+
Q_2\,s^2\right],
\ee
with
\ba
Q_1&=&-24+24\,\tau^2-2\,\tau^4+39\,g^2-17 \tau^2\,g^2-8 g^4,\\
Q_2&=&(-3+\tau^2+2\,g^2)/8.
\ea
From these expressions it is easy to compute the new distribution function
for the ellipticity,
\be
p_e^{\rm obs.}(e)=
p_e^{\rm prim.}(e)\left[1+\sigma_{\kappa}^2 {18 (-1+20\,e^2-16\,e^4)\over
(1+8\,e^2)^2}\right].\label{peobs}
\ee
Remarkably the term in $s^2$ disappears. The relation (\ref{peobs})
is one of the central results of this
paper. In the following 
this result is compared with results of numerical experiments.

\subsection{The extrema ellipticity distribution}

\subsubsection{For a Gaussian field}

The calculation of the ellipticity of the extrema has been examined
in detail by Bond \& Efstathiou (1987) in case of a 2D Gaussian field.
The principle of the calculation relies on the
computations of the number of peaks, $\Npk(\tau,\vg)$ of a given 
curvature (within the range $\d\tau$ and $\d^2\vg$). If the extrema have the
directions $\gamma_p$ in a given sample then,
\ba
&\Npk(\tau,\vg)\,\d\tau\,\d^2\vg=\nonumber\\
&\disp{\int\d^2\gam\sum_p\delta_{\rm Dirac}(\gam\!-\!\gam_p) \Prob(\tau,
\vg\vert \nabla\delta_T\!=\!0)\,\d\tau\,\d^2\vg},
\ea
where $\Prob(\tau,\vg\vert \nabla\delta_T\!=\!0)$ is the probability
distribution function of the local curvature under the constraint that
the local gradient vanishes.
In case of a Gaussian field this expression simplifies since
the local gradients are statistically independent of the curvature,
i.e.,
\be
\Prob^{\rm prim.}(\tau,\vg\vert \nabla\dTp\!=\!0)=\Prob^{\rm prim.}(\tau,\vg).
\ee
The trick to complete the calculation is then to express
the Dirac delta functions in $\gam-\gam_p$ in terms of the local curvature.
Indeed, for peaks we can Taylor expand the local
temperature gradient in,
\be
(\delta_T)_{,i}(\gam)=(\delta_T)_{,ji}(\gam_p)(\gam-\gam_p)_{j},
\ee
since, by definition, $(\delta_T)_{,i}(\gam_p)=0$.
As a result we have for the primordial field,
\be
\Npk(\tau,\vg)\,\d\tau\,\d^2\vg=N_{\rm tot.}\,
\disp{J(\tau,\vg)\,\Prob(\tau,\vg)\,\d\tau\,\d^2\vg\over
\int\,J(\tau,\vg)\,\Prob(\tau,\vg)\,\d\tau\,\d^2\vg},
\ee
where $N_{\rm tot.}$ is the total number of peaks, and $J(\tau,\vg)$
is the Jacobian of the transform between the positions and the
gradients, i.e.,
\be
J(\tau,\vg)=\vert Det(c_{ij})\vert=\vert\tau^2-\vg^2\vert.
\ee
As a result we have,
\be
\npk^{\rm prim.}(\tau,\vg)\ \d\tau\ \d^2\vg=
\disp{
\vert\tau^2-\vg^2\vert\,p^{\rm prim.}(\tau,\vg)\ \d\tau\ \d^2\vg,
\over \int\,
\vert\tau^2-\vg^2\vert\,p^{\rm prim.}(\tau,\vg)\ \d\tau\ \d^2\vg},
\ee
which, after simple calculations, implies that,
\be
p_{\rm peak.}^{\rm prim.}(e)\d e=
\disp{24\ \sqrt{3}\ (1-4\,e^2)\ e\,\d e \over (1+8\,e^2)^{5/2}},
\ee
when the integration is limited to the maxima (extrema with positive 
curvature), that is when $\tau^2> g^2$ and $\tau>0$.

\subsubsection{The peak ellipticity distribution with lens effects}

In this section I investigate the effects of lenses for the ellipticity
distribution of the extrema. The principle of the calculation
is very similar derivation to 
the Gaussian field case. The only change is that
in principle the local curvature may not be independent
of the gradient. Actually for peaks we know that (\ref{ACA}) is exact so that
we can express the curvature of the lensed map in terms of the
curvature of the primordial map for which the statistical properties are
completely defined. The effect of lenses will then show up
in the distribution function of the local curvature and in the expression
of the Jacobian. Taking these two changes into account we have,
\be
p^{\rm obs.}_{\rm peaks}(\tau,g_1,g_2)=
p^{\rm prim.}_{\rm peaks}(\tau,g_1,g_2)\left[1+Q_3\,\sigma_{\kappa}^2\right],
\ee
with
\be
Q_3=-28+24\,\tau^2+39\,g^2-2\,\tau^4-17\,\tau^2\,g^2-8\,g^4.
\ee
Note that in this case it is necessary to make sure that 
the integral is correctly normalized. The number density of peaks
remain unchanged with (sub-critical) lenses but the variances
for the curvature or the gradient are perturbatively affected.
From the previous result we have,
\be
p^{\rm obs.}_{\rm pk}(e)=
p^{\rm prim.}_{\rm pk}(e)\left[1+
\disp{2\,(-15+512\,e^2-112\,e^4)\over(1+8\,e^2)}\,
\sigma_{\kappa}^2\right].\label{pepeakobs}
\ee
A straightforward examination of (\ref{peobs}) and (\ref{pepeakobs}) 
show that the effect
of lens distortion is significantly larger for the peaks compared to the
field points. 

\subsection{The effects of filtering}

So far, in these analysis, I have ignored the effects of filtering.
Actually the CMB temperature maps will be observed with a finite
resolution. The measured temperature map is thus given by
\be
\tilde\delta_T(\gam)=
\int\d^2\gam'\,W(\vert\gam-\gam'\vert/\theta_0)\ \delta_T(\gam'),
\ee
where $W$ is a window function that in practice can be assumed to be
Gaussian. It is anyway possible to further filter the resulting
maps. In particular it can be very interesting to filter
the large angular scale modes out in order to reduce the cosmic variance.
In all cases, the filtered lensed temperature map is given by
\be
\tilde\delta_T^{\rm obs.}(\gam)=\int\d^2\gam'\,W(\vert\gam-\gam'\vert/\theta_0)
\delta_T^{\rm prim.}(\gam'+\xi(\gam')).
\ee
It is obviously not equivalent to the lens distortion effects
applied to the filtered primordial temperature map! The two operations do not
commute in general. This is a standard problem when one is considering
the effects of non-linear couplings. In this particular 
problem there is however a limiting case for which 
the problem can be completely solved. Indeed the CMB temperature
maps has an intrinsic small-scale cutoff due to the Silk
damping effect. If then the filtering scale is much smaller
than this intrinsic scale one can assume that locally 
$\delta_T^{\rm prim.}(\gam')\approx \delta_T^{\rm prim.}(\gam)$
when $\gam'$ is close enough to $\gam$. As a result we have
\ba
(\tilde\delta_T^{\rm obs.})_{,i}(\gam)\approx&
(\delta_T^{\rm prim.})_{,j}\,\times\nonumber\\
&\left[\delta^K_{ji}+\int\d^2\gam'\,W(\vert\gam-\gam'\vert/\theta_0)
\xi(\gam')_{j,i}\right].
\ea
The filtering effect applies only to the displacement field.
The same situation occurs for the 2nd order derivatives. In this limiting 
case, the effect of filtering is then simply to reduce the lens effect
by filtering the displacement field.
It implies in particular that the $\sigma_{\kappa}^2$ factor that intervenes
in the expressions (\ref{peobs}) and (\ref{pepeakobs}) is actually given by,
\be
\sigma_{\kappa}^2=\disp{{1\over 4}
\int{\d^2\vl\over (2\pi)^2}\,\p2d(l)\,W_{\theta_0}^2(l)}.\label{sigkf}
\ee
In practice however this approximation may not be very accurate. Even
at 4.5' arcmin resolution scale (the best channels of Planck surveyor)
about 20 to 25\% of the primordial signal is filtered out because
of the finite resolution. One problem that then should be addressed
with numerical experiment is the validity of this approximation.

\section{Numerical results}

The numerical results are based on a series of 20 realizations
of both the CMB temperature fluctuations and the displacement maps.
Each map is initially 10x10 $\deg^2$ large, with a pixel size of 1.17'
(these are 512x512 grid maps). After regridding to take into
account the lens effects and filtering, the actual useful size is
9x9 square degrees. In Figs. 2-4 I give the averaged ellipticity 
distributions obtained from these 20 realizations, 
and the error-bars correspond to the estimated cosmic
variance of one map.

\subsection{The ellipticity distribution}

\begin{figure}
\vspace{13 cm}
\special{hscale=60 vscale=60 voffset=-50 hoffset=-70 psfile=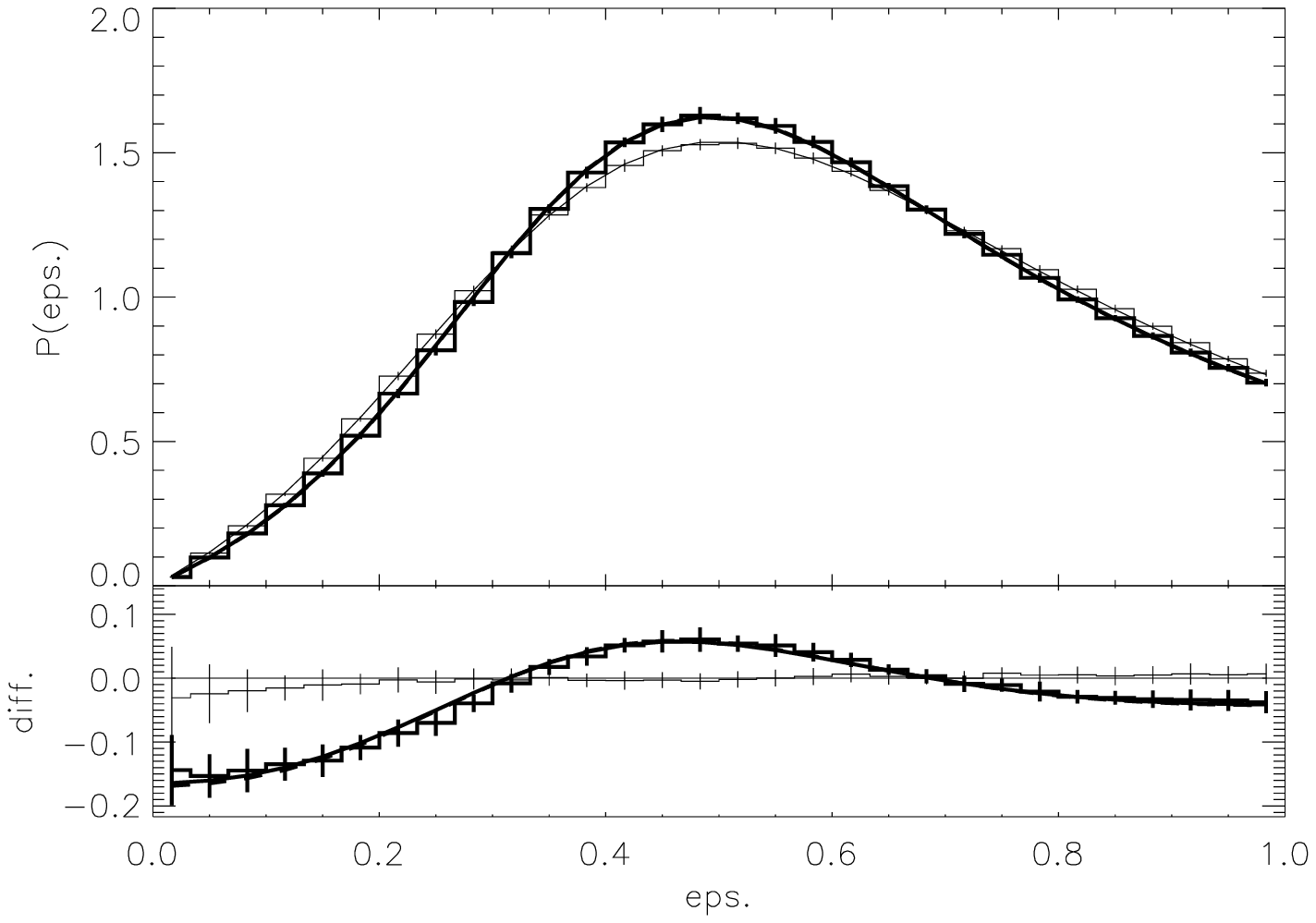}
\special{hscale=60 vscale=60 voffset=-235 hoffset=-70 psfile=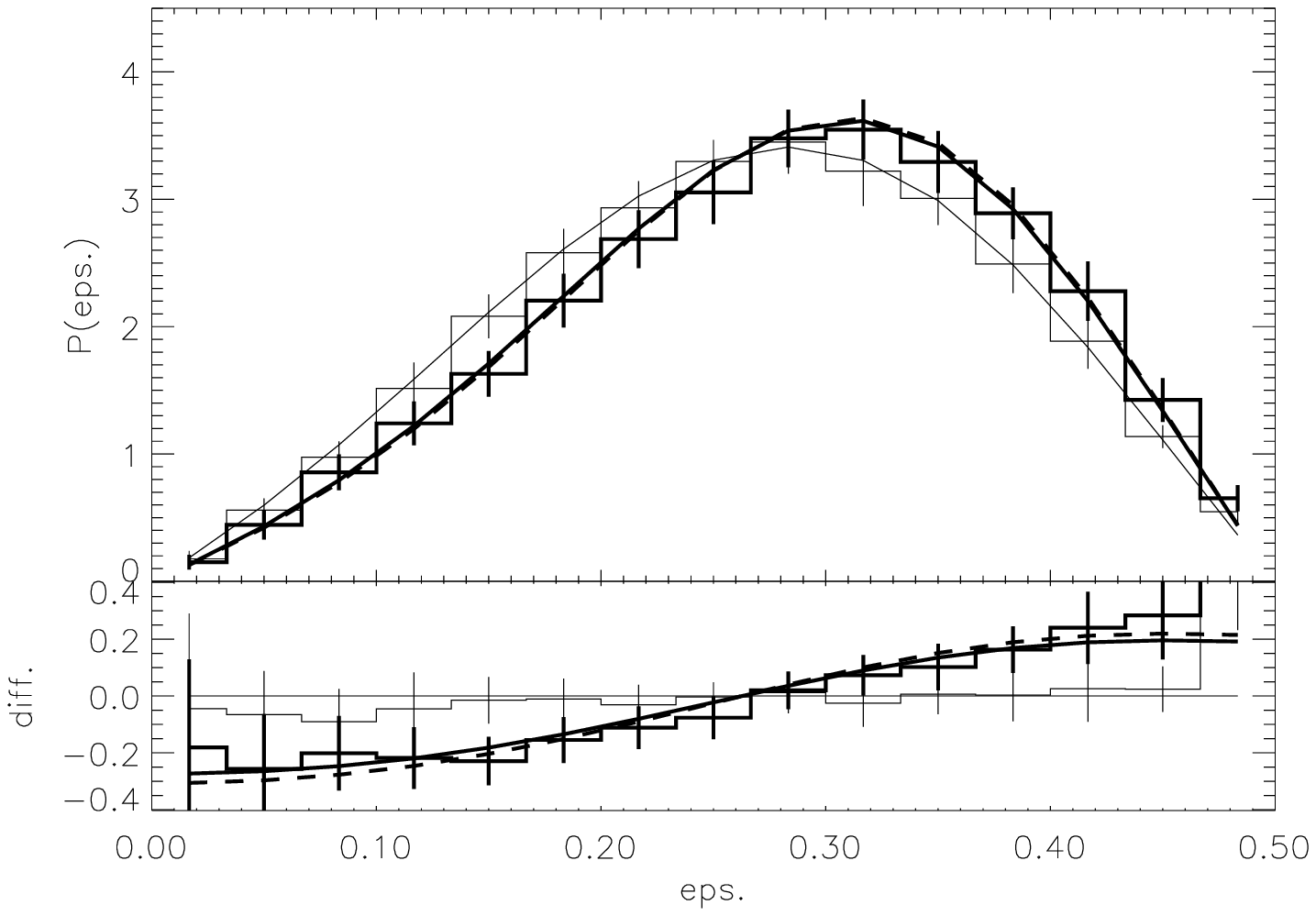}
\caption{The ellipticity statistics at 2.5 arcmin scale for $\sigma_8/0.6=2$.
The statistics for the field points is given in the top panels 
and for the extrema points in the bottom ones. The thin lines
correspond to the primordial temperature map, the thick lines to the
lensed map. 
The error bars correspond to the cosmic variance
computed for a 9x9 square degree size map. 
The cosmic variance has been estimated with
20 different realizations of both the primordial temperature maps and the 
displacement fields. The continuous lines give the theoretical predictions
(\ref{peobs}) and (\ref{pepeakobs}) with $\sigma_{\kappa}$ given by 
(\ref{sigkf}). The dashed lines correspond to the best fit of the
ellipticity distribution using formulae (\ref{peobs}) or 
(\ref{pepeakobs}) where $\sigma_{\kappa}$ is assumed to be a free parameter.
}
\end{figure}

\begin{figure}
\vspace{13 cm}
\special{hscale=60 vscale=60 voffset=-50 hoffset=-70 psfile=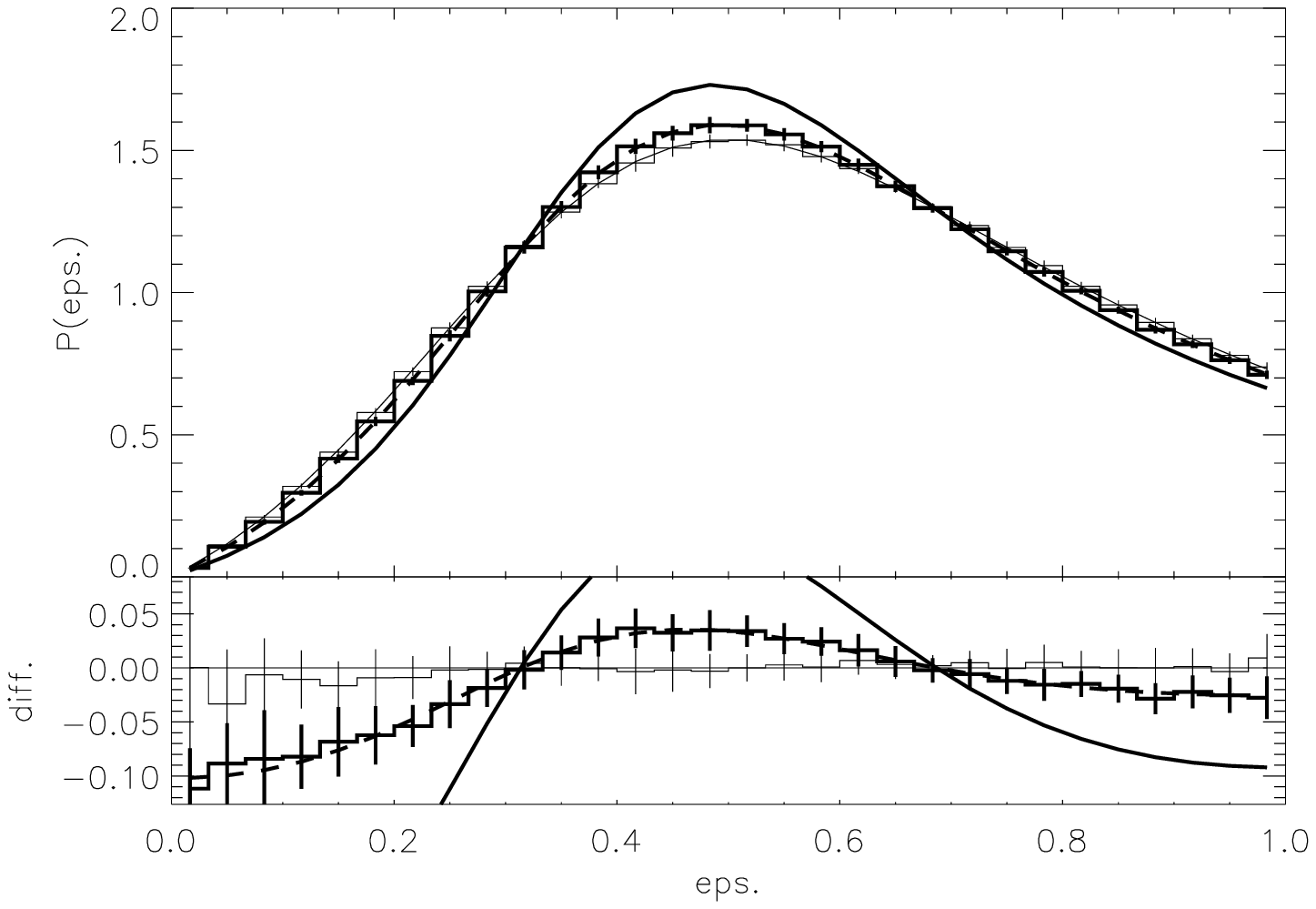}
\special{hscale=60 vscale=60 voffset=-235 hoffset=-70 psfile=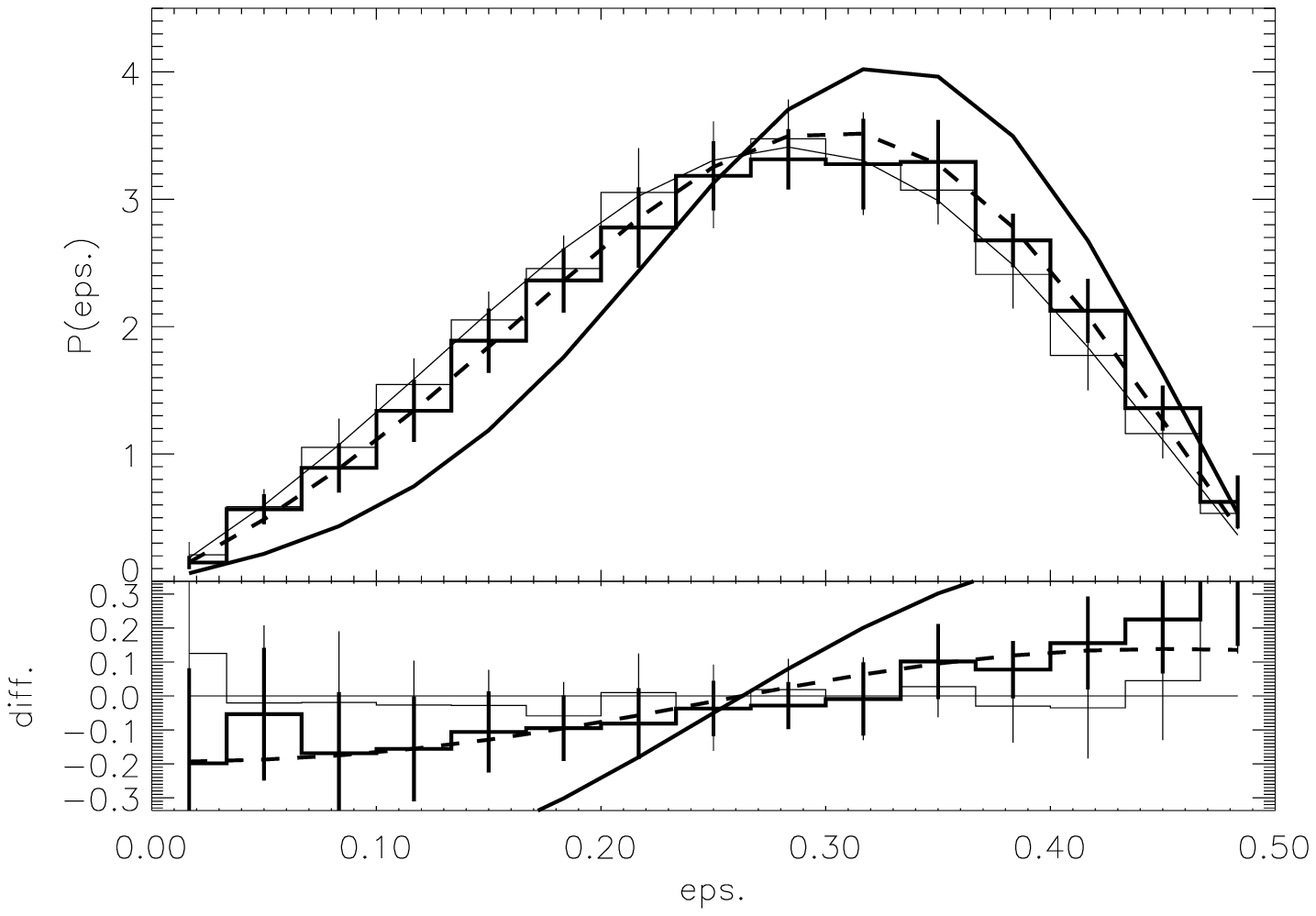}
\caption{Same as figure 2 with a 5 arcmin filtering scale, $\sigma_8/0.6=4$}
\end{figure}

The maps are filtered with a compensated
filter built as the difference of two Gaussian
window functions with one filtering radius being twice the other,
\be
W_{\theta_0}(l)=\exp(-l^2\,\theta_0^2/2)-
\exp(-2\,l^2\,\theta_0^2).
\ee
This make the maps rather weakly sensitive to the large angular scale
correlation. The correlation length of the local quantities
is thus expected to be rather small, of the order of $2\ \theta_0$.

The local curvature is calculated from finite differences. To be more
precise the scheme which is used is the following,
\ba
\delta_{T,11}(i,j)=&\disp{
{1\over 3}
\sum_{k=-1,1}\delta_T(i-1,j+k)-}\nonumber\\
&2\,\delta_T(i,j+k)+\delta_T(i+1,j+k);\\
\delta_{T,12}(i,j)=&\disp{{1\over 4}
\left[
\delta_T(i-1,j-1)-\delta_T(i-1,j+1)-\right.}\nonumber\\
&\disp{\left.\delta_T(i+1,j-1)+\delta_T(i+1,j+1)\right].}
\ea
where $\delta_T(i,j)$ is the local temperature fluctuations
at the grid point $(i,j)$.
The Figs 2-3 present the results of the distribution function
for various filtering scales, and various normalization of the
lens effects (expressed in terms of $\sigma_8$).
We can see that for a very small filtering scale, 2.5 arcmin, the
numerical results are 
in good agreement with the theoretical predictions
for both the field points and the extrema.

At larger angular, the filtering effect starts to play a major role and
significantly reduces the lens effects. 
Thus at 5 arcmin angular scale, the amplitude 
of the effect seems indeed to be reduced by a factor of about 2 compared
to what would be expected when the smoothing effects are neglected
(see table 1). This discrepancy is shown in figure 3 where
the dashed lines correspond to the best fit of the lens effect when
$\sigma_{\kappa}$ is a free parameter. The quality of the fit shows
however that the filtering effect changes the amplitude of the lens
effects, but not much the $e$  dependence of the corrective terms. 

\subsection{The Cosmic Variance}

Because of the use of compensated filters, it is fair
to assume that the a
whole sky survey would simply correspond
to a given number of {\it independent} $9\times 9$ square degree maps. 
This should be particularly accurate for the local statistical
indicators considered here. The cosmic variance
then scales as the square root of the number of such maps. 
The signal to noise ratio for the lens effect is thus expected to obey
the scaling property,
\be
\disp{
{\rm Signal \over Noise}\propto {\sigma_8^2\over\sqrt{\rm Sky\ coverage}}}.
\ee
If one is able to get a the sky coverage of 50\% as anticipated for the Planck
surveyor then it means that the cosmic variance should be, for the whole 
survey, about 16 times less important than for a $9\times 9$ square degree 
map.  The signal to noise ratio is expected to be the same
for a ``whole'' sky survey with $\sigma_8=0.6$ or for a $9\times 9$ 
square degree map with $\sigma_8=4\times 0.6$. The latter case then provides
us with a realistic estimation of the Cosmic Variance
for the future Planck mission.

\begin{table*}
\caption{Values of $\sigma_{\kappa}$ from direct determination with the
displacement maps (eq. \ref{sigkf})
and from best fit parameterizations. Errors give
the cosmic variance for maps of 9x9 square degree size.}
\begin{tabular}{ccccccc}
&&&
$\sigma_{\kappa}^2\times\,10^{3}$, best fit,&$\sigma_{\kappa}^2\times\,10^{3}$, best fit,&
$\sigma_{\kappa}^2\times\,10^{3}$, best fit,&$\sigma_{\kappa}^2\times\,10^{3}$, best fit,\\
&&$\sigma_{\kappa}^2\times\,10^{3}$&field points, &field points, & extrema,&extrema,\\
$\sigma_8$&ang. scale&direct det.&no lens effects&with lens effects&no lens effects&with lens effects\\
\hline
\vspace{-.2cm}
&&&&&\\
0.6&2.5'&$2.3\pm0.1$&$0.2\pm1.2$&$2.5\pm1.4$&$0.2\pm1.8$&$4.0\pm1.7$\\
1.2&2.5'&$9.2\pm0.3$&$0.2\pm1.2$&$9.4\pm1.5$&$1.2\pm1.8$&$10.3\pm1.8$\\
1.2&5'&$5.1\pm0.3$&$0.4\pm1.6$&$1.8\pm1.6$&$1.0\pm2.1$&$2.5\pm2.1$\\
2.4&5'&$20.\pm1.$&$0.4\pm1.6$&$5.6\pm1.6$&$1.0\pm2.1$&$6.5\pm2.2$\\
2.4&5' + noise&$20.\pm1.$&$0.4\pm1.5$&$5.6\pm1.6$&$1.2\pm2.0$&$6.5\pm2.1$
\end{tabular}
\end{table*}

In table 1, I summarize the results obtained for the determination
of $\sigma_{\kappa}$ with a best fit and the error associated
with this determination. This error has been determined as the r.m.s.
of the determined $\sigma_{\kappa}^2$ among the 20 realizations at my
disposal. One can see that the signal ratio grows indeed with
$\sigma_8$ and scales like $\sigma_8^2$ for a given
filtering scale. The filtering effects however
reduce the signal by a factor 3 from what
would be expected with a direct determination of $\sigma_{\kappa}$.
The noise is independent on $\sigma_8$
but depends slightly on the filtering scale. 
The signal to noise ratio expected for the Planck survey
at 5' resolution scale is thus about 3 if $\sigma_8=0.6$.

\subsection{The effect of Noise}

\begin{figure}
\vspace{6. cm}
\special{hscale=60 vscale=60 voffset=-235 hoffset=-70 psfile=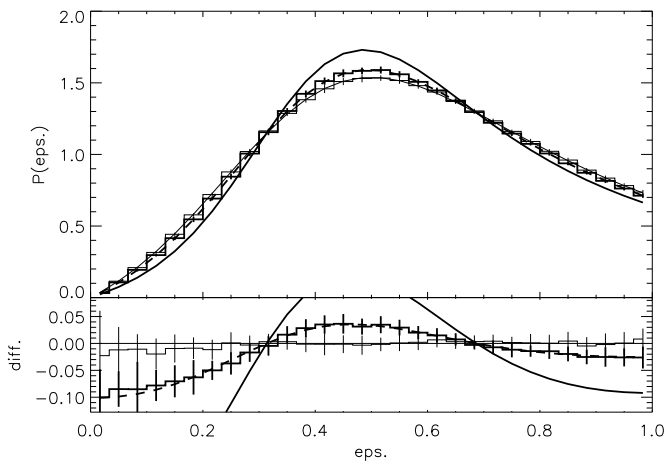}
\caption{Same as figure 2 with a 5 arcmin filtering scale, $\sigma_8/0.6=4$
and including noise}
\end{figure}

Fig. 4 and the last line of the table show the effects of noise
of the measured map.
Its effect is found to be very weak. However the noise here has been
assumed to be a pure white noise, and thus, for the
ellipticity statistics, it reproduces the Gaussian case. 
This is certainly not a realistic assumption. 
The striping introduces indeed nonlocal correlations
that may affect significantly statistics that are related to local topological
quantities. The study of the striping effect is very dependent
on the experiments and is beyond the scope of this paper.

\section{Discussion}

\subsection{The dependence on the cosmological parameters}

\begin{figure}
\vspace{6 cm}
\special{hscale=50 vscale=50 voffset=-185 hoffset=-30 psfile=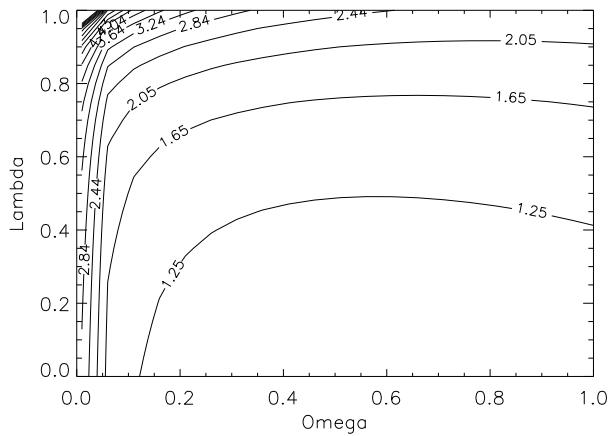}
\caption{The ratio $\sigma_{\kappa}^2(\Omega,\Lambda) \Omega^{-1}/
\sigma_{\kappa}^2(\Omega=1,\Lambda=0)$ as a function
of $\Omega$ and $\Lambda$ for a power law spectrum with $n=-1.5$.}
\end{figure}

The dependence on the cosmological parameters is explored here in a quite
rough way. The temperature power spectrum, and its relation to
the density power spectrum, has indeed a rather
complicated dependence on all the cosmological parameters.
So in this simple study I will assume that the shapes
of $C_l$ and $P(k)$ remain the same and simply discuss the dependence
of the amplitude of 
$\sigma_{\kappa}$ on the cosmological parameters due to 
the density-convergence relationship.
The ratio $\sigma_{\kappa}/\sigma_8$ is obviously a quantity
that one might want to consider but it should be compared
to other constraints coming from large-scale structure formation.
The number density of clusters can provide us with a
constraint which is in principle free of 
bias contamination. It constrains the
amplitude of the density fluctuations at roughly the $8\,h^{-1}$
Mpc scale for an Einstein-de Sitter Universe (Oukbir et al 1997), 
which corresponds roughly to the scale of interest for the lens
CMB effects. For open universes the dependence cannot be given simply
in terms of $\sigma_8$ because linear mass scale of the galaxy clusters
is shifted to a larger scale. A dependence arises then with the
slope of the power spectrum (Oukbir \& Blanchard 1997). Ignoring these
subtleties, and following Eke, Cole \& Frenk (1996)
I will simply assume that the actual constraint coming from the observed
number density of rich clusters can be written,
\be
\sigma_8\ \Omega^{0.5}\approx 0.6\pm 0.1.
\ee
It means that the larger $\Omega$ the smaller $\sigma_8$.
To estimate the variation of $\sigma_{\kappa}$
with roughly a fixed number density of clusters, one should then
compute $\sigma_{\kappa}^2(\Omega,\Lambda)\ \Omega^{-1}/
\sigma_{\kappa}^2(\Omega\!=\!1,\Lambda\!=\!0)$. The result is given in
Fig. 5. One can see, as expected, that the magnitude of the 
signal is very weakly $\Omega$ dependent if $\Lambda=0$
but significantly grows with $\Lambda$. It implies that
the signal to noise ratio would be a factor about 2 larger
for a $\Lambda$-CDM.

\subsection{On the interest of lens effect detection.}

The observational quantities that have been investigated here have
been designed to be sensitive to the non-linear couplings induced by
lens effects.  It is of course possible to consider other quantities,
such as the Minkowski functionals, whose general properties for a
Gaussian field have been recently investigated in detail (Winitzki \&
Kosowski 1997 and Schmalzing \& G\'orski 1997).  The lens effects
might indeed significantly affect the Minkowski functional for
2D maps. For instance the shapes of high threshold peaks
are shown to be more elongated, which implies that the averaged
circumference near  the top of the peak should increase.   However, a
complete theoretical investigation of this effect is quite difficult
because this is a nonlocal indicator.  And in general, the most
efficient non-Gaussian indicators probably depend on the processes one
wants to detect. The number of hot and cold spots seems to be a good
way to detect topological defects.  Is is obviously not the case for
the lenses.

In  a previous paper the four-point function
was considered. Actually the local curvature can be viewed
as an other way to have access to the same information. Here
the four point function is simply averaged in a way 
that avoid too much cancellation. And other possible
quantity is the collapsed four-point function, that identifies with
the four order cumulant of the local temperature. This quantity
however tends to cancel the contribution of the various terms.
This seems not to be the case for the local curvature.
It does not mean however that we cannot do better 
for detecting the lens effects. For instance the two-point 
correlation function of the local ellipticity
(with orientation) might be a good indicator. 
However one should have in mind
that the number of CMB structures per lens is rather small.
It is thus not obvious that it can improve the situation. 

Note that the results obtained for the lens effects, temperature
four-point function or ellipticity statistics, are all sensitive to the 
lens two-point correlation function.
It comes from the fact that the only quantity related to the
lenses which is potentially available from CMB maps is its power-spectrum.
The non-Gaussian properties of the lens population are
for instance not accessible.

If it is actually possible to detect a lens effect,  the smallness
of the signal to noise ratio with which it can be determined
indicates  that it will be
pointless to use this information as a mean to constrain
the cosmological parameters. But, it would be
extremely interesting to be able to check that the amount of lens effects
computed from models favored by the CMB power spectrum are in agreement
with its detection, or its non-detection.

Note however that
other secondary effects might as well induce non-Gaussian properties.
The Rees-Sciama effect in particular is due to the evolving
non-linear potentials and has a source term which is intrinsically
non-Gaussian. A peculiar case of this effect is the moving lens
effect (Birkinshaw \& Gull 1983): 
potential wells that move perpendicularly to the
line of sight induce temperature fluctuations. This effect is
quadratic with the cosmic fields (proportional to the local
velocity times the gradient of the potential) and is thus intrinsically
non-Gaussian. These cases are probably worth investigating. I expect however
that they give a smaller non-Gaussian signal since the static
lens effect is the only mechanism that couples the primary anisotropies
to the line of sight potentials. 

\section*{Acknowledgments}

Part of this work has been initially pursued with 
Thibaut Marrel who is warmly thanked for it.
The author is also grateful to Y. Mellier for
many fruitful discussions and a careful reading of the manuscript, 
to U. Seljak and B. Jain for the use of their codes.



\begin{thebibliography}{}

\bibitem{} Barnes, C. \& Turok, N. 1997, hep-ph/9702377
\bibitem{} Bernardeau, F. 1997, A\&A, 324, 1
\bibitem{} Birkinshaw, M. \& Gull, S.F. 1983, Nature, 302, 315
\bibitem{} Blanchard, A. \& Schneider, J. 1987, A\&A, 184, 1
\bibitem{} Bond, J.R., Efstathiou, G. 1987, MNRAS, 226, 655 
\bibitem{} Cay\'on, L., Mart\'\i nez-Gonz\'alez, E. \& Sanz, J.L. 1993a, ApJ, 403, 471
\bibitem{} Cay\'on, L., Mart\'\i nez-Gonz\'alez, E. \& Sanz, J.L. 1993b, ApJ, 413, 10
\bibitem{} Cole, S., Efstathiou, G. 1989, MNRAS, 239, 195
\bibitem{} Fukushige, T., Makino, J. \& Ebisuzaki, T. 1994, ApJ, 436, L107
\bibitem{} Hamilton, A.J.S., Matthews, A., Kumar, P. \& Lu E. 1991, ApJ, 374, 1
\bibitem{} Jain, B., Mo, H. J. \& White, S.D.M. 1995, MNRAS, 276, L25 
\bibitem{} Jungman, G., Kamionkowski, M., Kosowsky, A. \& 
Spergel, D.N., 1996, astro-ph/9512139
\bibitem{} Kashlinsky, A. 1988, ApJ, 331, L1
\bibitem{} Linder, V.E., 1990, MNRAS, 243, 353
\bibitem{} Oukbir, J., Blanchard, A. 1997, A\&A, 317, 1
\bibitem{} Oukbir, J., Bartlett, J.G., Blanchard, A. 1997, A\&A, 320, 365
\bibitem{} Peacock, J.A. \& dodds, S. J.1996, MNRAS, 282, 877
\bibitem{} Pen, U., Spergel, D., Turok, N. 1994, Phys. Rev. D, 49, 692
\bibitem{} Sasaki, M. 1989, MNRAS, 240, 415
\bibitem{} Smoot, G. F. et al. 1993, ApJ, 396, L1
\bibitem{} Seljak, U. 1996, ApJ 463, 1.
\bibitem{} Seljak, U. \&  Zaldarriaga, M. 1996, ApJ, 469, 437
\bibitem{} Schmalzing, J. \& G\'orski, K.M. 1997, astro-ph/9710185
\bibitem{} Suginohara, M., Suginohara, T., Spergel D. N. 1997, 
astro-ph/9705134 
\bibitem{} Tomita, K. \& Watanabe, K. 1989, Prog. Theor. Phys., 82, 563
\bibitem{} Turok, N. 1996, ApJ, 473, L5
\bibitem{} Winitzki, S. \& Kosowsky, A. 1997, astro-ph/9710164

\end{thebibliography}
\end{document}